\documentclass{article}
\usepackage{graphicx} % Required for inserting images

\title{Abundance vs Diversity}
\author{Marco Antonio Rosas Pulido}
\usepackage[a4paper, margin=1in]{geometry}
\usepackage{titlesec}
\usepackage{lipsum} % Para texto de ejemplo
\usepackage[style=nature,backend=biber]{biblatex}
\usepackage{csquotes}
\usepackage{graphicx} % Paquete para manejar imágenes
\usepackage{xcolor}
\usepackage{float}
\usepackage{amssymb} % For \lesssim
\usepackage{authblk}
\usepackage{booktabs}
\usepackage{amsmath}
\usepackage[bottom]{footmisc}

\titleformat{\section}{\large\bfseries}{\thesection.}{0.5em}{}
\titleformat{\subsection}{\normalsize\bfseries}{\thesubsection.}{0.5em}{}

\title{Abundance and  Economic diversity as a descriptor of cities' economic complexity}

\author{Marco A. Rosas\textsuperscript{1}, 
        Roberto Murcio\textsuperscript{2},
        Omar R. Vázquez \textsuperscript{1},
        Carlos Gershenson\textsuperscript{3}}

\affil{\small\textsuperscript{1}C3, UNAM, Mexico. \quad
       \textsuperscript{2}Birkbeck, University of London, UK. \quad
       \textsuperscript{3}SUNY Binghampton, USA.}

\affil{\small\texttt{marco.rosas@c3.unam.mx}, 
       \texttt{r.murciovillanueva@bbk.ac.uk}, 
       \texttt{cgg@binghamton.edu}}

\begin{document}

\maketitle

\begin{abstract}
Intricate interactions among firms, institutions, and spatial structures shape urban economic systems. In this study, we propose a framework based on three structural dimensions — abundance, diversity, and longevity (ADL) of economic units — as proxies of urban economic complexity and resilience. Using a decade of georeferenced firm-level data from Mexico City, we analyze the relationships among ADL variables using regression, spatial correlation, and time-series clustering. Our results reveal non-linear dynamics across urban space, with power-law behavior in central zones and logarithmic saturation in peripheral areas, suggesting differentiated growth regimes. Notably, firm longevity modulates the relationship between abundance and diversity, particularly in peri-urban transition zones. These spatial patterns point to an emerging polycentric restructuring within a traditionally monocentric metropolis. By integrating economic complexity theory with spatial analysis, our approach provides a scalable method to assess the adaptive capacity of urban economies. This has implications for understanding informality, designing inclusive urban policies, and navigating structural transitions in rapidly urbanizing regions.
\end{abstract}

\section{Introduction}

Urban sustainability involves not only human–environment relations, but also economic and social interactions that shape the organization and evolution of contemporary societies \parencite{Gershenson2013}. Many urban models conceptualize cities primarily through the spatial allocation of economic and demographic functions. Yet, there is an increasing effort to integrate these frameworks with representations of urban morphology, which has been shown to display characteristic fractal properties \parencite{Batty2009}. Urban dynamics can be described by integrating four dimensions: economic, social, environmental, and institutional. When coupled, these variables create a system with a high complexity \parencite{Spangenberg2005}. In this context, economic units—particularly organizations such as firms, government agencies, labor unions, and political parties—are key components of modern economic and social systems. Their structure and behavior have significant implications for systemic resilience and transformation \parencite{Hannan1992}. It is well established that economic structural transformation plays a crucial role in sustainable development. However, existing literature has largely focused on internal changes within specific economic sectors, often overlooking external relationships. This raises the question of the role that external factors play in structural economic sustainability transformations at different scales \parencite{Truong2021}.

From the perspective of institutional economics, the study of these organizations can be approached through different paradigms. While new institutional economics tends to focus on individual-level analysis, classical and heterodox institutionalism emphasize collective dynamics and systemic configurations \parencite{Angela2024}. This distinction provides a conceptual lens for examining organizational structures and their relationship to certain properties of spatial distribution, such as the abundance and diversity of economic units. In this sense, the spatial patterns of economic activities can be interpreted not only as outcomes of market dynamics but also as expressions of institutional arrangements and collective behaviors embedded within specific territorial contexts.

Despite the availability of extensive quantitative data, relevant knowledge gaps persist regarding the life cycle of economic units. Firms may disappear through mergers, splits, or liquidation, yet the determinants of such processes—such as age, size, or type of activity—are not fully understood. Although considerable attention has been given to firm size distributions, less is known about the distribution of firm lifespans \parencite{Daepp2015}. These lifespans are shaped by a wide range of interactions among economic agents, which may take the form of competition, conflict, negotiation, or cooperation \parencite{Bartolini1999}. Such interactions are closely linked to the structural properties of the economic environment in which the units operate. In this work, we review two main structural properties: the number of units (\textit{abundance}) and the types of units (\textit{diversity}) found in a particular geographic area.

Regional growth and resilience are increasingly viewed as outcomes of these structural features, rather than the mere presence of specific sectors. Studies have shown that relatedness and complexity metrics can capture these structural features without making strong assumptions about their nature \parencite{Neffke2011, Cottineau2019}, opening novel interpretations about how spatial economic structures shape development trajectories.

In recent years, economic complexity frameworks have emerged as powerful alternatives to traditional models of economic growth. Unlike classical approaches that rely on aggregate outputs (e.g., GDP) or predefined factors (e.g., labor, capital), complexity methods leverage detailed datasets of thousands of economic activities and their spatial distribution to infer latent structures and interaction patterns \parencite{Hidalgo2009, Hidalgo2021, Balland2020, Hao2021}. Particularly relevant are the dimensions of \textbf{abundance, diversity, and longevity (ADL)}, which may serve as indicators of systems' adaptive capacity and underlying complexity.

Overall, the link between urban spatial structure and economic performance remains murky in both the theoretical and empirical literature. In part, the problem stems from shortcomings in defining the spatial scale of analysis and urban form \parencite{Zhang2017}.

The spatial distribution of economic diversity plays a critical role in shaping regional resilience to external shocks. As \parencite{Shen2025} and \parencite{Li2023} note, related variety functions as a shock diffuser, mitigating impacts through the risk-spreading effect, while unrelated variety serves as a shock absorber with significant direct effects and non-negligible spatial spillovers. Similarly, \parencite{Li2023} found that regions with high related variety exhibit strong economic resilience in the post-crisis period, whereas unrelated variety shows no significant direct influence on long-term recovery. However, both forms of variety generate local spatial spillovers that enhance resilience over the next year following the 2008–2009 crisis, even if these effects diminish over longer horizons. Taken together, these findings confirm that industrial relatedness and the composition of neighboring economies are central to short-term recovery capacity, underscoring the importance of spatially interconnected economic diversity in fostering urban and regional robustness.

%\subsection{Research Problem}

In this work, we explore the relationships among abundance, diversity, and Longevity of economic units in Mexico City and propose that these structural features can serve as measures of economic complexity and resilience. Special emphasis is placed on identifying potential \textit{non-linear relationships} among these variables and understanding their implications for the adaptive performance of economic systems. We aim to contribute to a deeper understanding of how economic units survive and adapt within competitive markets. Furthermore, by incorporating a spatial econometric perspective, it becomes possible to assess the influence of regional proximity, technological diffusion, and agglomeration effects—factors that may generate spillover benefits for less complex but geographically connected economies \parencite{Fingleton2000, Rey2001, Lundberg2006}. This interdisciplinary approach enhances the analysis of urban economic resilience through the lens of complexity science and spatial data analytics.
%In particular, it is unclear whether these structural dimensions correlate with the resilience and transformation capacity of local economic systems. In the Mexican context, the marked heterogeneity among states raises further questions about how local economic structures influence firm survival and regional performance \parencite{Chavez2017}.

%\subsection{Objective}

%This research aims to examine whether the Abundance diversity, and longevity of economic units can serve as descriptors of economic complexity and resilience. Special emphasis is placed on identifying potential \textit{non-linear relationships} among these variables and understanding their implications for the adaptive performance of economic systems.

%\subsection{Relevance}

%The economic performance of a metropolitan area the size of Mexico City is a crucial indicator of its economic health. Changes in the spatial organization of the central city will inevitably affect the functioning of the economy in metropolitan areas, as it is the epicenter of the region's growth \parencite{Shen2025}. However, only a few empirical studies have examined the effect of changes in the physical layout of central cities on the economic growth of metropolitan areas.

\section{Methods}

\begin{figure}[H]
    \centering
    \includegraphics[width=1\textwidth]{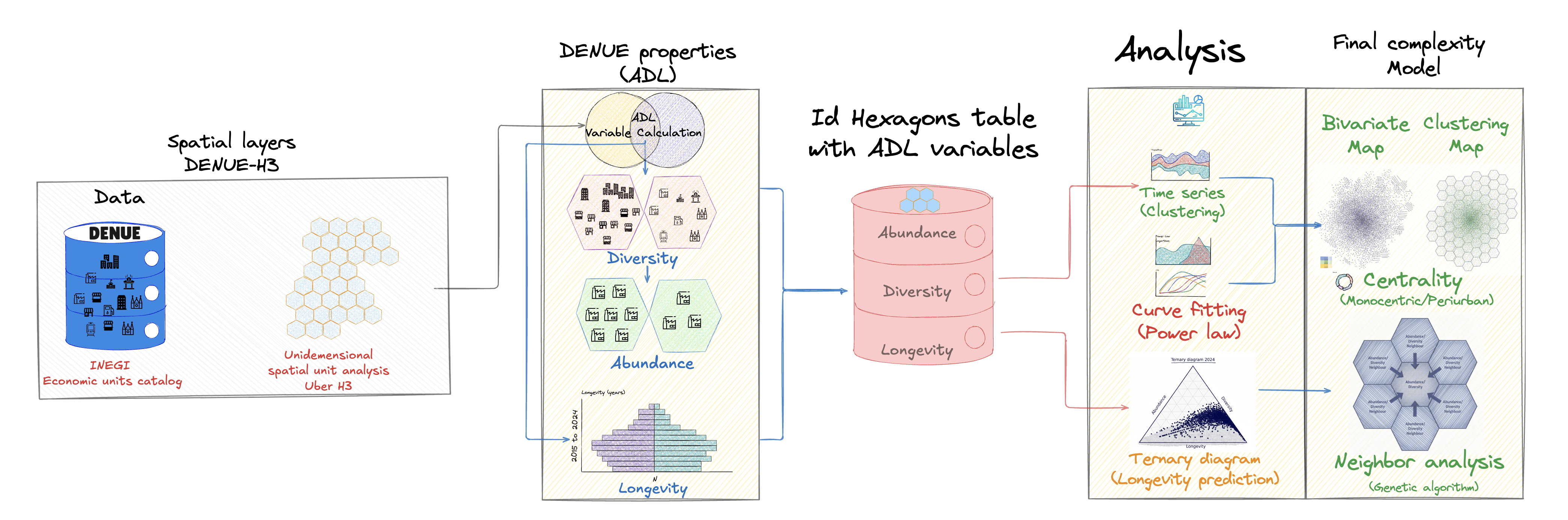}
    \caption{Workflow for the spatial and statistical analysis of firm-level data in Mexico City. The diagram summarizes the integration of geospatial and statistical procedures, beginning with combining the DENUE (National Statistical Directory of Economic Units) point layer with the Uber H3 hexagonal grid to generate polygonal spatial layers for analysis. 
    Using INEGI’s georeferenced DENUE catalog of economic units, three properties—abundance, diversity, and longevity (ADL)—were derived for each economic unit. These properties were calculated based on the composition of the units within a homogeneous or one-dimensional spatial analysis unit (H3) for each of the ten years analyzed.
    Subsequently, these three properties were aggregated for each hexagon: (1) abundance, defined as the total number of economic units within the hexagon; (2) diversity, defined as the number of distinct economic activities present; and (3) longevity, calculated as the average number of years each economic unit appears within the same hexagon.
    Using this information, a database was constructed in which each record corresponds to a uniquely identified hexagon in Mexico City and contains the three variables representing the ADL properties.}
    %At the hexagonal level, three structural variables are calculated: Abundance (total number of firms), Diversity (heterogeneity of economic activities), and Longevity (persistence of firms between 2015 and 2024). These variables serve as the basis for subsequent analyses, including time-series clustering, spatial correlation through centrality mapping to identify monocentric behavior and peri-urban patterns, power-law curve fitting to capture scaling behavior, and finally, neighbor analysis, which will be proven by using a genetic algorithm to evaluate the prediction of spatial interdependencies to longevity prediction. Together, these steps will provide a comprehensive framework for examining the spatial organization and temporal evolution of urban economic systems.}
    \label{img:framework}
\end{figure}

Agglomeration economies depend not only on the presence of economic units but also on the spatial configuration of their interactions \parencite{Alcacer2014, Cottineau2019, Duranton2004, Eberts1999, Puga2010}. Measuring the multidimensional nature of urban economic interactions requires a spatially explicit approach that accounts for both the distribution and density of economic activity. 
In this study, we propose a methodological framework (Figure \ref{img:framework}) to capture the multidimensional nature of intra-city economic interactions, analyzing both \textit{dimensional} and \textit{spatial} relationships of a diverse pool of economic units over 10 years.

Using a decade of geo-referenced firm-level data from Mexico City, combined with a hexagonal spatial index (H3), we conducted spatial and statistical analyses to examine the relationships among three structural variables: \textbf{abundance}, \textbf{diversity}, and \textbf{longevity} (ADL). These relationships are explored through time-series clustering, regression analyses and spatial correlation measures, providing a comprehensive assessment of the structural dynamics of urban economic activity.

\begin{itemize}
    \item \textbf{Time Series clustering}
    To explore the temporal evolution of abundance and diversity variables at the hexagon level, we applied the Dynamic Time Warping algorithm to cluster hexagons based on the temporal behaviour of these variables.  
    \item \textbf{Dimensional relationships:} Linear correlation and regression analyses were used to quantify the strength and direction of associations between ADL variables.
    \item \textbf{Spatial relationships:} Spatial-neighbor analysis was performed to examine whether the ADL characteristics of a hexagon’s neighbors influence its ADL profile. The longevity variable, calculated as the number of consecutive years a hexagon maintained activity, as described in Ecuation:\eqref{eq:equation4} and included in the ADL dataset, was used as the response variable in the model. While the abundance–diversity correlation within each hexagon did not predict longevity, the abundance–diversity characteristics of neighboring hexagons significantly predicted it. Spatial autocorrelation techniques, such as Moran’s I and Local Indicators of Spatial Association (LISA) \parencite{Chen-Moran2023}, were employed to detect patterns of clustering or dispersion.
    
    %Spatial-neighbor analysis was performed to examine whether the ADL characteristics of a hexagon's neighbors influence its ADL profile. Spatial autocorrelation techniques, such as Moran's I and Local Indicators of Spatial Association (LISA), were employed to detect patterns of clustering or dispersion .
\end{itemize}

\subsection{Data}
Economic unit data were gathered from the Mexican National Statistical Directory of Economic Units (DENUE). This dataset provides updated identification and location data for active economic units across the national territory, for the planning, design, and evaluation of public economic policies. It is also a reliable source of statistical information and a fundamental part of the decision-making process for direct investment, optimizing resources in both the public and private spheres.

\begin{figure}[H]
    \centering    \includegraphics[width=0.45\textwidth]{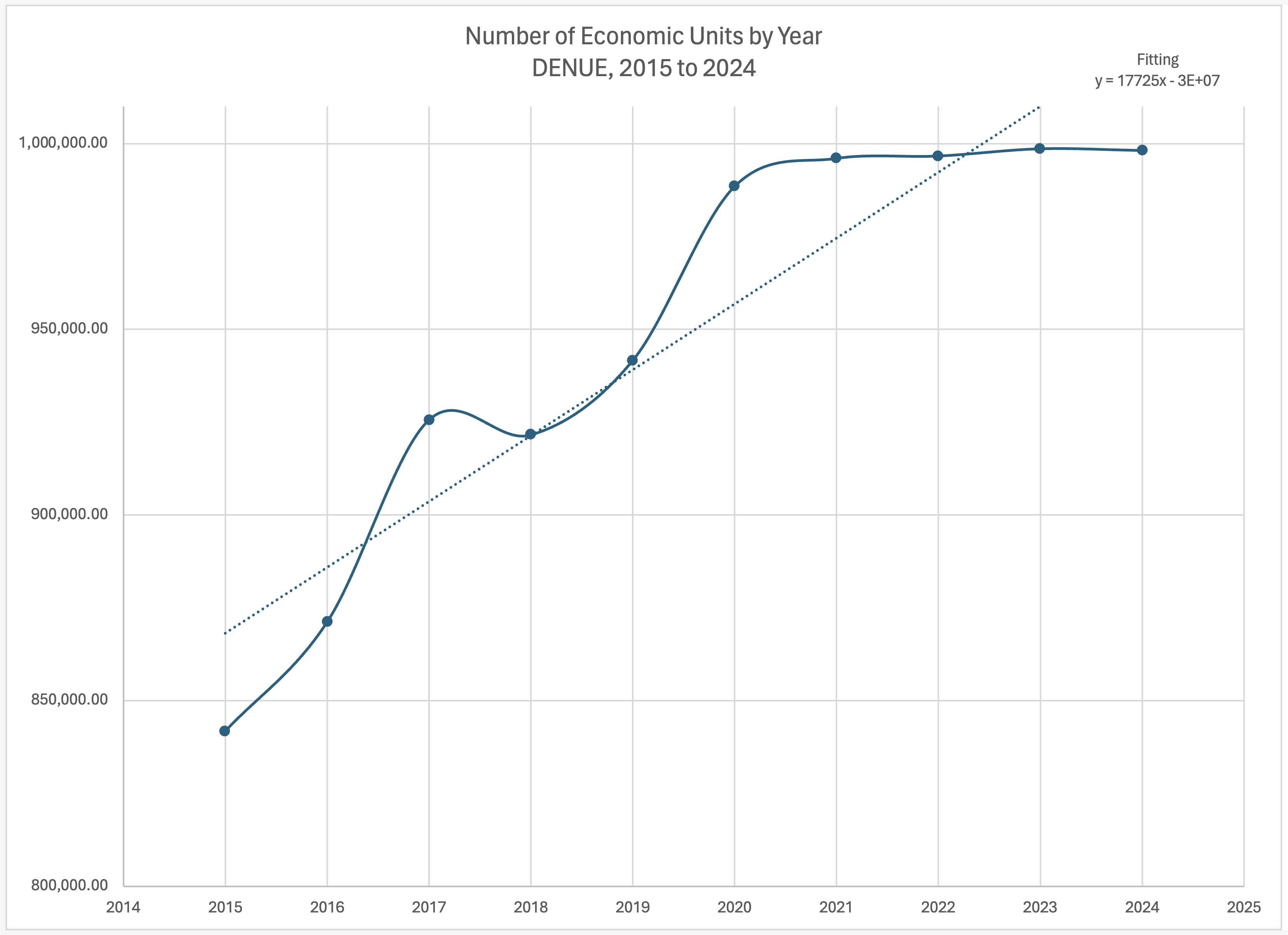}
\includegraphics[width=0.45\textwidth]{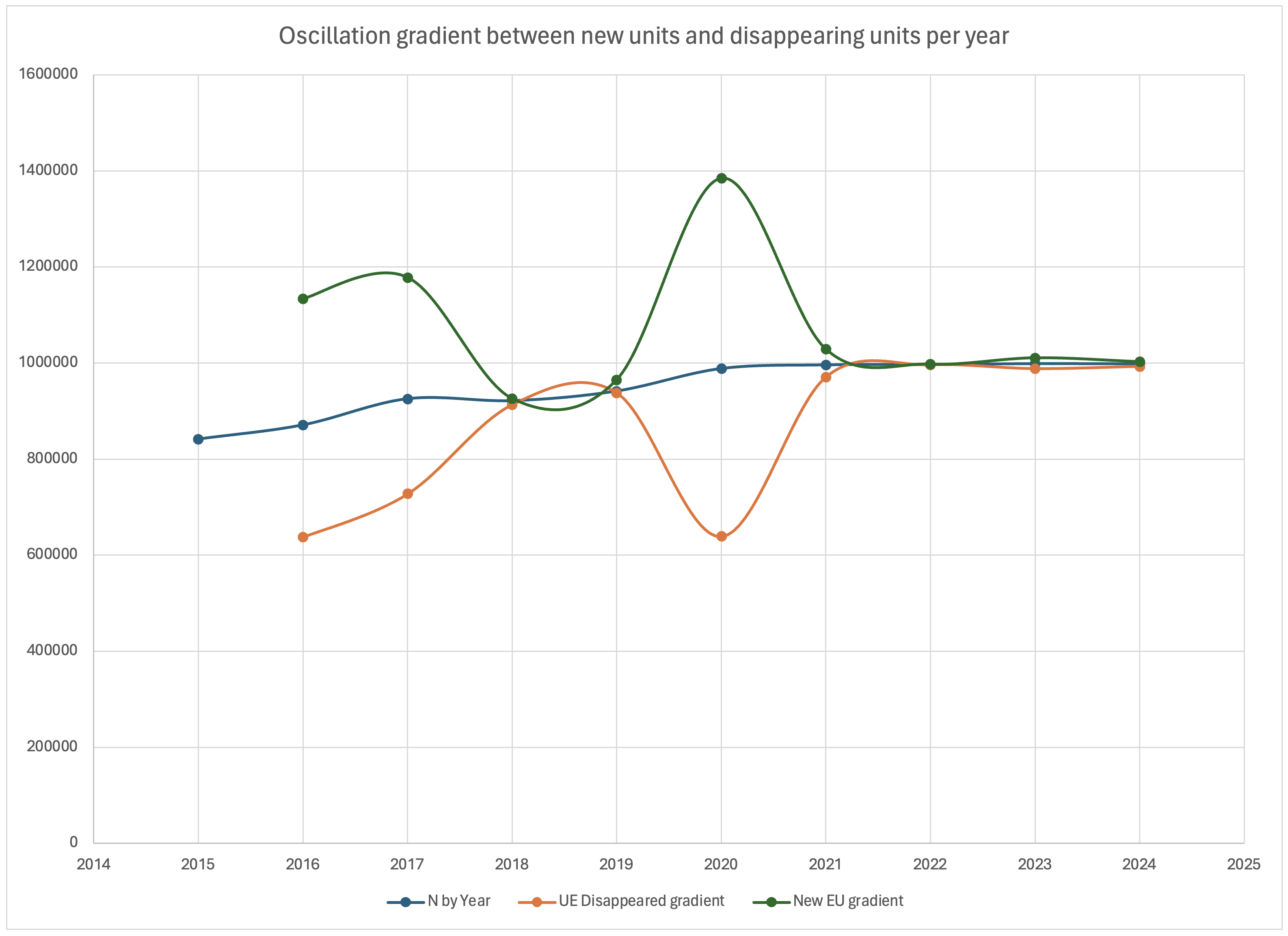}
    \caption{The total number of units registered DENUE (INEGI) between 2015 and 2024 (left). We can observe a clear upward trend, with a peak in 2017, followed by a period of lower variation beginning in 2019. (right) New economic units registered vs those that disappeared from the directory each year .
}
    \label{fig:nunits}
\end{figure}

This dataset comprises yearly information from 2015 to 2024, where each economic unit is represented by a point (latitude, longitude) layer, totalling 9,479,882 records and averaging 947,988 per year. The full structure of the DENUE data is described in Appendix A. The number of economic units per year (Figure \ref{fig:nunits}, left) is not constant, depicting a net growth of 156,334 units over the 10 years studied.  (Figure \ref{fig:nunits}, right) shows the oscillation between the new and disappearing units, with near-equilibrium points in 2018 and 2019. Notably, a sharp increase in new economic units in 2017 explains the atypical behavior observed that year. From 2020 onward, both lines tend to stabilize.

%The economic unit that is defined as a single physical location, permanently established in a place and delimited by fixed buildings and installations, combines actions and resources under the control of a single owner or controlling entity to carry out some economic activity, whether for profit or not. Includes dwellings in which economic activities are carried out. This means that in a single location, several economies can coexist. 

The total number of distinct commercial establishments operating betweenacross the different versions of the Directory published and as the API access key Technical Standard for the incorporation and updating of information in the Statistical Registry of Businesses of Mexico [],  the Business Statistical Key (CLEE) was incorporated into the Directory table, whose structure allows for the precise and unique identification of each establishment and company registered in the Statistical Registry of Businesses of Mexico (RENEM) and the DENUE.
%To maintain the comparability of information between the different versions of the Directory, the publication of the DENUE ID key is maintained to identify each establishment. This key serves as a binding factor for the data between the different versions of the Directory published and as the access key to the API \parencite{DENUE_INEGI}.

To capture the spatial distribution of economic units while maintaining geometric consistency and minimizing edge effects, we adopted the H3 hierarchical spatial index developed by Uber. H3 is a system of tessellated hexagons that allows for efficient, scalable, and hierarchical spatial analysis \parencite{brodsky2018h3}. Unlike conventional square grids, hexagons provide equal-distance neighbors, avoid directional bias, and facilitate smoother neighborhood calculations, which are essential for capturing spatial interactions.
We defined a hexagonal grid with a radius of 600 meters (1.2 km) that covers the Mexico City metropolitan area. This resulted in a total of 9,644 hexagonal spatial units. The resolution was chosen to balance the granularity needed to observe local variations while maintaining computational feasibility and consistency across space.

%\begin{itemize}
 %   \item \textbf{Point layer:} Economic units from the National Statistical Directory of Economic Units (DENUE) provided by INEGI \parencite{DENUE_INEGI}, which contains detailed georeferenced information on businesses operating in Mexico.
%    \item \textbf{Polygon layer:} The H3-based hexagonal grid, used as the analytical spatial unit for aggregation and comparison.
%\end{itemize}

%Each economic unit from the DENUE database was spatially assigned to a corresponding hexagon. The time frame considered for the analysis spans between 2015 and 2024, allowing the construction of a temporal profile for each hexagonal cell.
\subsection{Time Series Clustering}
Once the abundance and diversity measures were calculated, we constructed a 14-point time series (one point per year from 2010 to 2024) for each $H_i$. These time series reflect the annual evolution of the local economic structure, allowing us to track temporal changes in abundance and diversity.
Although each $H_i$ possesses a unique temporal trajectory, we hypothesize that within the study area, certain hexagons may exhibit structurally similar temporal profiles, i.e., sequences with similar shapes, indicating parallel patterns of economic dynamics. These similarities may reflect common trajectories of economic change, such as synchronized growth, decline, or stabilization, potentially driven by shared structural or contextual factors (e.g., location).
To test this hypothesis, we followed a time series clustering approach based on the Dynamic Time Warping (DTW) algorithm \cite{Yadav2019}, which is a robust method for comparing time series that may be temporally misaligned, vary in amplitude, or evolve at different speeds. This is particularly useful in our approach, where changes in abundance and diversity do not necessarily occur at the same pace across different hexagons. To implement the DTW, we used the dtwclust package in R \cite{sarda2019time}

By clustering the $3,988$ hexagons based on their DTW similarity, we aim to uncover latent spatio-temporal groupings that reflect common development pathways. Identifying these temporal clusters helps to gain a better understanding of how local economies evolve under shared pressures or opportunities. It may offer insights into the emergence of economic regimes or structural tipping points within the metropolitan system.
\subsection{Abundance, diversity, and longevity (ADL) definition}

The ADL measures are defined for each hexagon $H_i, i={1,2,3,...,n}$ and for each economic unit $p_j, j={1,2,3,...,n}$
\subsubsection{Abundance}
The total number of economic units within each hexagon for each year is defined as 

\begin{equation} \label{eq:equation1}
%\[
A_{H_i} = \sum_{i=1}^{n} \mathbf{1}_{\mathcal{H_i}}(p_j)
%\]
\end{equation}
where $A_{H_i}$ is the abundance  \( \mathbf{1}_{\mathcal{H_i}}(p_j) \) is an indicator function equal to 1 if the economic unit \( p_i \) lies within the hexagon \( \mathcal{H_i} \), and 0 otherwise.

\subsubsection{Diversity}
Number of distinct economic activity codes (based on NAICS \footnote{NAICS (North American Industry Classification System) is a standardized system established in 1997 by the United States, Canada, and Mexico to classify businesses and industries. It assigns numeric codes that group economic activities into sectors and subsectors, facilitating the collection, analysis, and comparison of economic data across North America.} classification) within each hexagon:

\begin{equation} \label{eq:equation2}
D_{H_i} = \left| \left\{ c_i \mid p_j \in \mathcal{H_i} \right\} \right|
\end{equation}
where $D_{H_i}$ is the total number of distinct activities associated with economic units contained in \( \mathcal{H_i} \) and \( c_i \) represents the economic activity code of economic unit \( p_j \). 

%Once the hexagon table containing these two variables was created, we began by analyzing the correlation between Abundance and diversity per spatial unit of analysis (hexagon). 

\subsubsection{Longevity}
Number of years each economic unit appears in the DENUE dataset, regardless of the year it appeared or ceased to be present:
\begin{equation} \label{eq:equation3}
\mathrm{L}_{p_j} = t_{p_j}^{\mathrm{end}} - t_{p_j}^{\mathrm{start}}+1.
\end{equation}
where $t_{p_j}^{\mathrm{start}}$ and $t_{p_j}^{\mathrm{end}}$ denote the first and last year that an economic unit $p_j$ is registered in the database.  
%We used this data to calculate the average mortality rate of economic units within a hexagon. 
\subsubsection{Average longevity per hexagon}
By aggregating at ${L}_{p_j}$ at $H_i$ level we obtain average lifespans (equation \ref{eq:equation4}). 

\begin{equation} \label{eq:equation4}
\overline{\mathrm{L}}_{h_i} = \frac{1}{A_{h_i}}\sum_{i\in n}\mathrm{L}_i
= \frac{1}{A_{h_i}}\sum_{i\in n}\big(t_{p_j}^{\mathrm{end}}-t_{p_j}^{\mathrm{start}}+1\big).
\end{equation}

$L_{h_i}$ is the average longevity of economic units in $h_i$, measured as function of their aggregated lifespan. 
It is essential to note that, to incorporate the longevity variable into the correlation analysis, the data were presented for 2024, as this is the most recent year available in the DENUE records. However, all previous years were included in the calculation of average longevity per hexagon. Although some economic units ceased operations before 2024, they were not excluded from the final analysis of the relationship among longevity, diversity, and abundance (Fig. 14). This approach accounts for longevity computations within each hexagon, including the operational duration of economic units that disappeared before 2024. For example, if an establishment operated from 2015 to 2019, those five years contribute to the average longevity value of the corresponding hexagon.

%\subsection{Data}

%To calculate the Abundance the economic units were counted for each hexagon polygon, grouping them by the corresponding polygon I

\section{Results}

\subsection{Time Series Analysis}
\begin{figure} [H]
    \centering
    \includegraphics[width=1\textwidth]{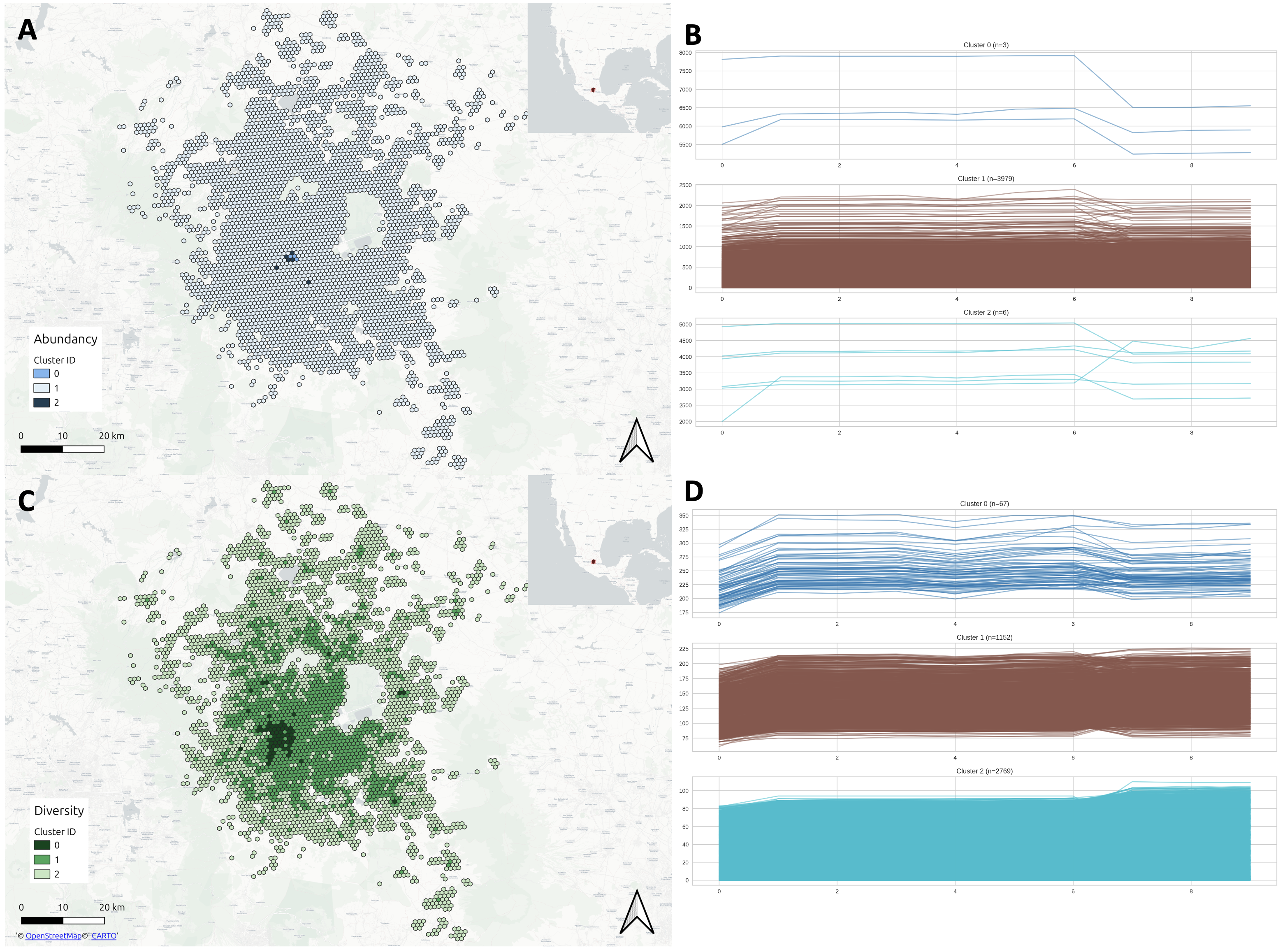} % 
    \caption{Time Series clusters for abundance (Figure B) and diversity (Figure D). In Figure A, $H_i$ diversity DTW clusters. The stable diversity throughout the city is evident. -say something about the three clusters in the middle. And, in Figure B, the three obtained clusters for the abundance time series. The monocentric nature of Mexico City is depicted in the concentric pattern observed.  }
    \label{img:TSclusters}
\end{figure}

 The DTW algorithm identified three main clusters in the abundance and diversity structural variables (Figure \ref{img:TSclusters}), suggesting that the temporal evolution of these structural variables follows a clear, homogeneous pattern in Mexico City. The behaviour of the abundance is practically the same for all hexagons (all but 9 are in cluster 1). On the other hand, diversity exhibits a slightly diverse behaviour, with cluster 1 accounting for 28.8\% and cluster 2 for 69.8\% of $3,988$ hexagons. 

%\begin{figure}[H] 
%   \centering
%    \includegraphics[width=0.9\textwidth]{Imagenes/cluster_abundanceKmeans.png} % 
%   \caption{$H_i$ diversity DTW clusters. The stable diversity throughout the city is evident. -say something about the three clusters in the middle} 
%    \label{dtw}
%\end{figure}

%\begin{figure}[H] 
%    \centering
%   \includegraphics[width=0.9\textwidth]{Imagenes/cluster_diversityKmeans.png} % 
%   \caption{The three obtained clusters for the abundance time series. The monocentric nature of Mexico City is depicted in the concentric pattern observed. Clusters form a concentric pattern.... }
%   \label{dtw}
%\end{figure}

\subsection{Dimensional relationships}

\begin{figure}[H]'%''h' indica que la imagen debe colocarse aproximadamente aquí
    \centering
    \includegraphics[width=0.9\textwidth]{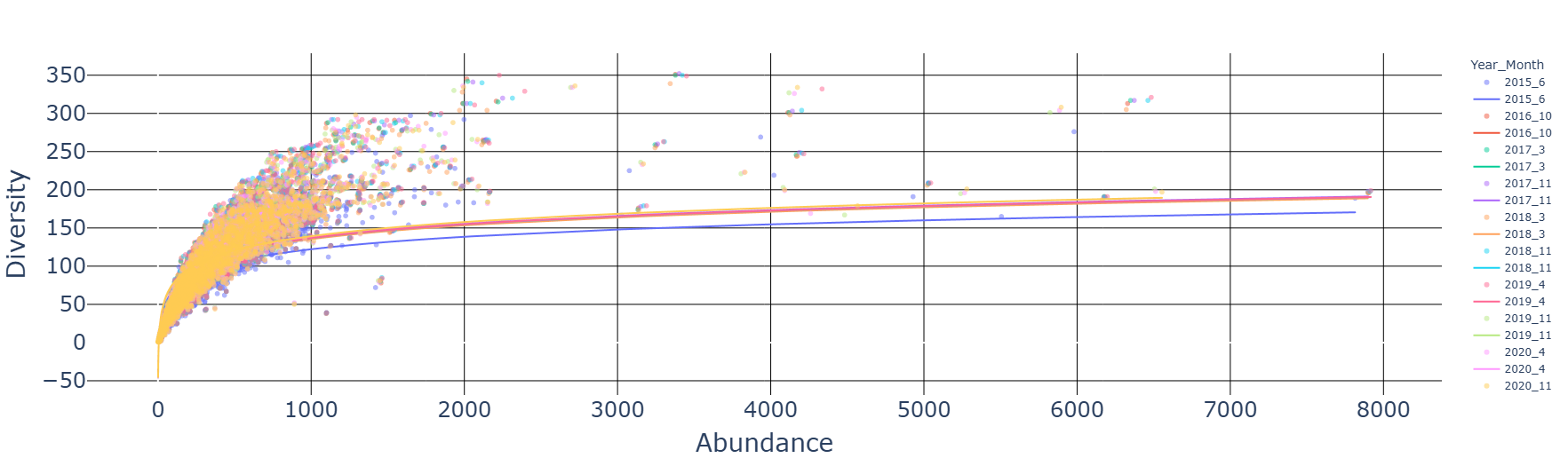} 
    \caption{Correlation between the abundance and diversity variables, all periods. Each point represents an $H_i$, and each line represents the obtained fit with the: -Ordinary least squares (OLS), Statmodels Python function used for the fit.
    Although a global non-linear trend is observed, three regimes, based on the abundance, are also identified, ranging from 1) $0$ to $1000$, 2) $1000$ to $2000$, and 3) from abundance $>2000$.}
    %Still lots to explain, like the diversity threshold at 350. Abundance is crazy high at the outliers: 8000!! What is in this hexagon? We need to locate it in space and say what is in it. Now that I'm thinking, we need to make it crystal clear the area of these hexagons, as abundances greater than 1000 are suspicious without a proper explanation.}
    \label{fig:scatterAD}
\end{figure}

First, we explored whether there is a statistical correlation between abundance and diversity during the studied period. The mean Spearman coefficient for the studied period is $\rho = 0.985$, indicating a strong relationship between these two variables. Moreover, figure \ref{fig:scatterAD} suggests that the correlation is non-linear or that more than one regime is needed to describe the interaction between the structural variables. In the following section, we explore both possibilities.

\begin{figure}[H]
    \centering
    \includegraphics[width=0.9\textwidth]{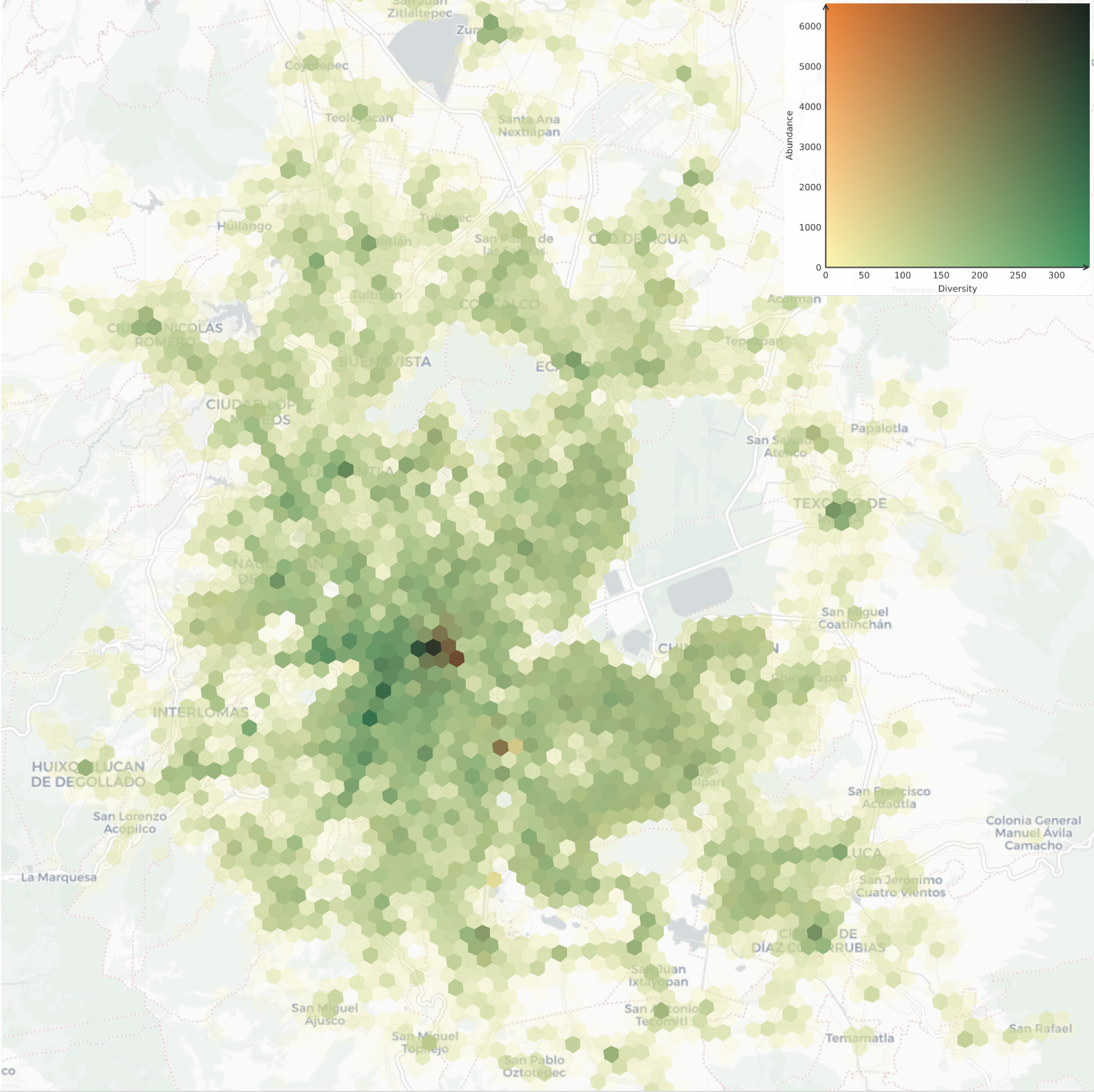}
    \caption{Mexico City abundance-diversity bivariate map. Explain something about the monocentric and ring structure observed. Also, what is in the North red Hexagons?}
    \label{figure:bivariate}
\end{figure}

\subsubsection{Correlation -regression- models}

To explore the complex relationship between abundance and diversity, we applied and compared linear and non-linear (power-law) regression models from 2015 to 2024. Table \ref{table:modelerrors} shows that the power law model consistently outperforms the linear model in terms of Root Mean Square Error (RMSE) and the coefficient of determination (R²). In 2015, the RMSE decreased from 14.58 to 13.94, and the R² increased from 0.62 to 0.68. In 2024, RMSE decreased from 14.69 to 14.22, while R² improved from 0.59 to 0.66. These results support the hypothesis that the relationship between abundance and diversity is non-linear and governed by scale-dependent dynamics, particularly in areas of low abundance.

\begin{table}[H]
\centering
\label{tab:comparison}
\begin{tabular}{@{}cccc@{}}

\toprule
\textbf{Year} & \textbf{Model} & \textbf{RMSE} & \textbf{R²} \\
\midrule
2015 & Linear     & 14.58 & 0.62 \\
2015 & Power Law  & 13.94 & 0.68 \\
2024 & Linear     & 14.69 & 0.59 \\
2024 & Power Law  & 14.22 & 0.66 \\
\bottomrule
\end{tabular}
\caption{Comparison between linear and power law regression models. All other years between 2015 and 2024 are shown in the Appendix section}
\label{table:modelerrors}
\end{table}

%Although the non-linear regression models explain almost $>70\%$ of the observed behavior, Figure \ref{fig:scatterAD} suggests that the relationship between abundance and diversity follows different regimes, based on the abundance value. 

\begin{figure}[H]
    \centering
    \includegraphics[width=0.85\textwidth]{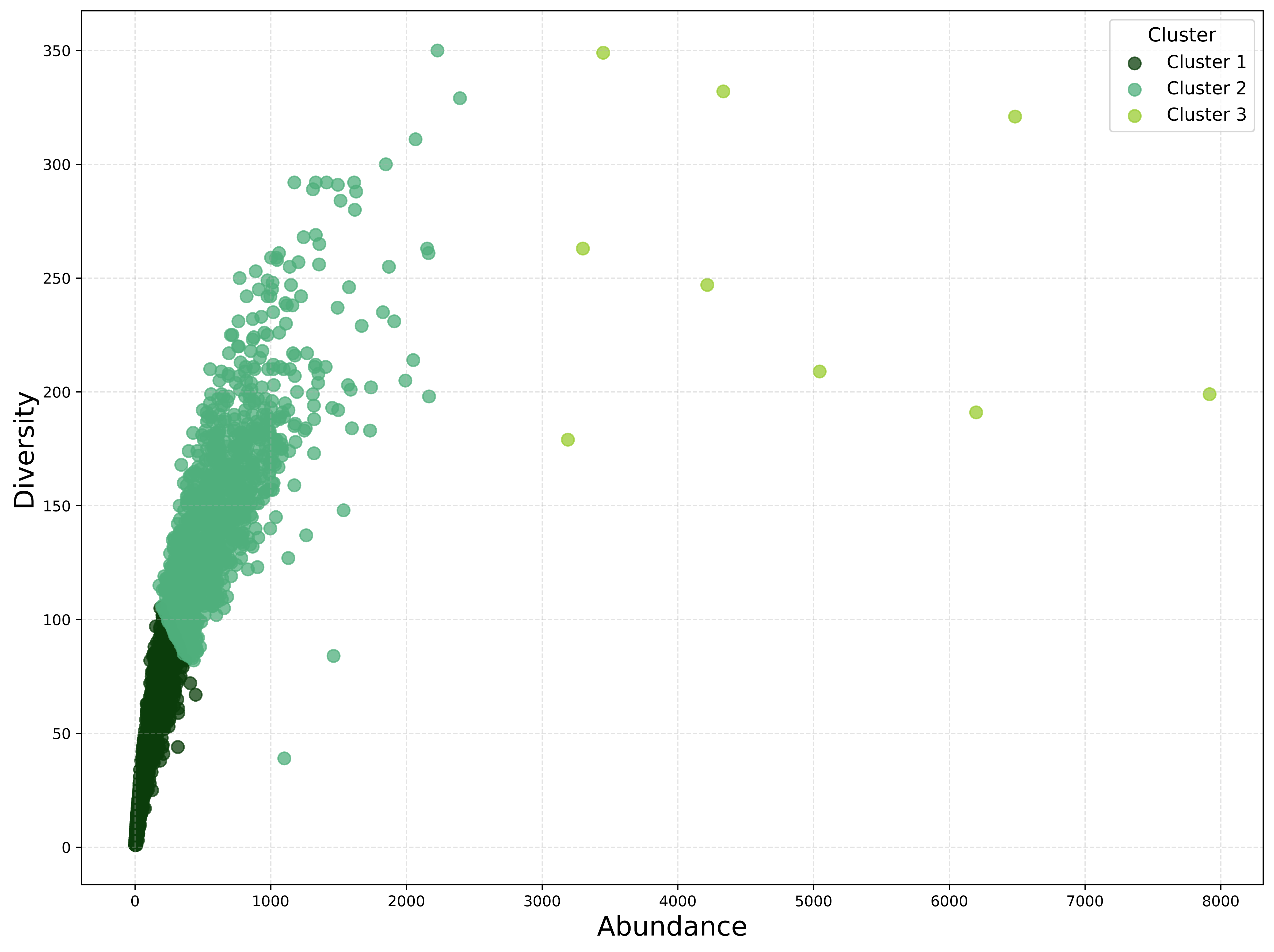}
    \caption{Obtained clusters for  2024. Clusters were defined using the elbow method.}
    \label{fig:clustersElbow}
\end{figure}

Exploring in more detail Figure \ref{fig:scatterAD}, we observed a clear shift after abundance~$1000$, where the increasing trend in diversity plateaus, oscillating between $150$ and $350$. 
To test this hypothesis, we stratified the data into the three clusters shown in Figure \ref{fig:clustersElbow}. The number of cluster we obtained using a traditional Elbow method. 

Subsequently, different fitting models were tested within each cluster for each year.
Non-linear models were fitted using least-squares optimization
(\texttt{scipy.optimize.curve\_fit}, and model performance was evaluated using RMSE; exponential models were systematically excluded due to fitting failure or divergent errors. Across the full period analyzed (2015--2024), Cluster~1 is consistently best described by a power-law relationship, with a highly stable scaling exponent ($0.55 \lesssim - \lesssim 0.57$) and minimal interannual RMSE variation, indicating a persistent scale-invariant structure with no evidence of regime shifts. In contrast, Cluster~2 is better captured by logarithmic or low-order polynomial forms, reflecting bounded, saturating dynamics with diminishing returns rather than scale-free behavior.

%\begin{figure}[h]
%    \centering
%    \includegraphics[width=0.45\textwidth]         {Imagenes/AD_2015_1000-2000 (1).jpg}
    %\includegraphics[width=0.45\textwidth]{Imagenes/AD_2024_1000-2000 (1).jpg}
    %\caption{Best fitting (logarithmic) for cluster two (1000-2000) in 2015 and 2024. (Note that the image does not accurately represent the three clusters, as they have different values.)}
%\end{figure}

%\begin{table}[htbp]
  
%    \begin{tabular}{|c|c|}
%\hline
%        Year & $\rho$   \\ \hline
%        2015 &  0.982 \\ \hline
%        2016 &  0.984 \\ \hline
%        2017 &  0.985 \\ \hline
%        2018 &  0.985 \\ \hline
%        2019 &  0.986 \\ \hline
%        2020 &  0.985 \\ \hline
%        2021 &  0.985 \\ \hline
%        2022 &  0.985 \\ \hline
%        2023 &  0.985 \\ \hline
%        2024 &  0.985 \\ \hline
%    \end{tabular}
%    \caption{Spearman rank correlation $\rho$. We observe high values across all years, indicating a strong non-linear correlation.}
%    \label{table:Spearman}
%  \hfill

\begin{table}[ht]
\centering
\begin{tabular}{|c|c|c|c|}
\hline
        Cluster & Fitting & RMSE & $R^2$\\ \hline
        1 & Power Law & 14.2168 & R \\ \hline
        2 & Logarithmic & 15.8124 & R \\ \hline
        3 & NA & NA & NA \\
         \hline
    \end{tabular}
    \caption{Best fit per cluster, 2024. For cluster 1 the best-fitted curve equation was 
    $y=4.48x^{0.55}-7.62$}
    \label{table:clusterFit}
\end{table}

%Initially (low abundance), diversity grows rapidly → few individuals, but of different species. Then (medium abundance), diversity reaches a maximum → equilibrium between the number of individuals and the number of classes. Finally (high abundance), diversity tends to decrease → a few classes dominate, displacing others. This pattern resembles a negative quadratic relationship (inverted parabola).

The analysis reveals a heterogeneous relation: 
Cluster one (abundance from $0$ to $1000$) exhibits power-law behavior, indicating the presence of scale-invariant, potentially self-organizing processes that amplify variability within the system \parencite{Cyert2020}. We hypothesize that this is because there are three orders of magnitude more small economic units than large ones. This reflects the predominance of many small firms alongside a few large ones—the latter being crucial to the overall functioning of the economic system. Empirically, the presence of a power-law distribution in firm size suggests that economic activity is concentrated among a relatively small subset of firms \parencite{Gabaix2016}. 

Cluster two (abundance from $1000$ to $2000$) displays logarithmic behavior, indicating a system that stabilizes or diversifies more evenly, suggesting diminishing marginal effects and a tendency toward stabilization as diversity increases \parencite{Vesperoni2023, Haisch2015}.

Cluster three comprises a relatively small number of observations with a wide dispersion, and although it is insufficient for robust statistical inference, it may represent a transitional regime or a boundary condition. 

This functional divergence across clusters underscores the need to consider context-specific mechanisms when analyzing the adaptive dynamics of economic/urban systems \parencite{Stanley1996, Batty2009}. Clusters represent not only structural but also functional groupings, revealing a clear spatial differentiation in the distribution of economic activities within the urban system. The central cluster, located in the historical and functional core of the city, exhibits power-law characteristics—marked by high concentration and inequality in both the density and diversity of economic units. This pattern indicates the presence of economic agglomeration processes, in which cumulative preference mechanisms reinforce centrality: activities tend to cluster around highly connected nodes, generating an internal functional hierarchy.
In contrast, the peripheral cluster, situated around the urban center, follows a logarithmic pattern, reflecting more homogeneous growth constrained by saturation effects. This configuration reveals a process of extensive urban expansion, where the increase in the abundance of economic units does not translate linearly into a proportional rise in diversity—likely due to structural constraints such as limited infrastructure, weak connectivity, or narrow functional specialization.

\subsubsection{Longevity analysis}
\begin{figure}[H]
    \centering
    \includegraphics[width=0.85\textwidth]{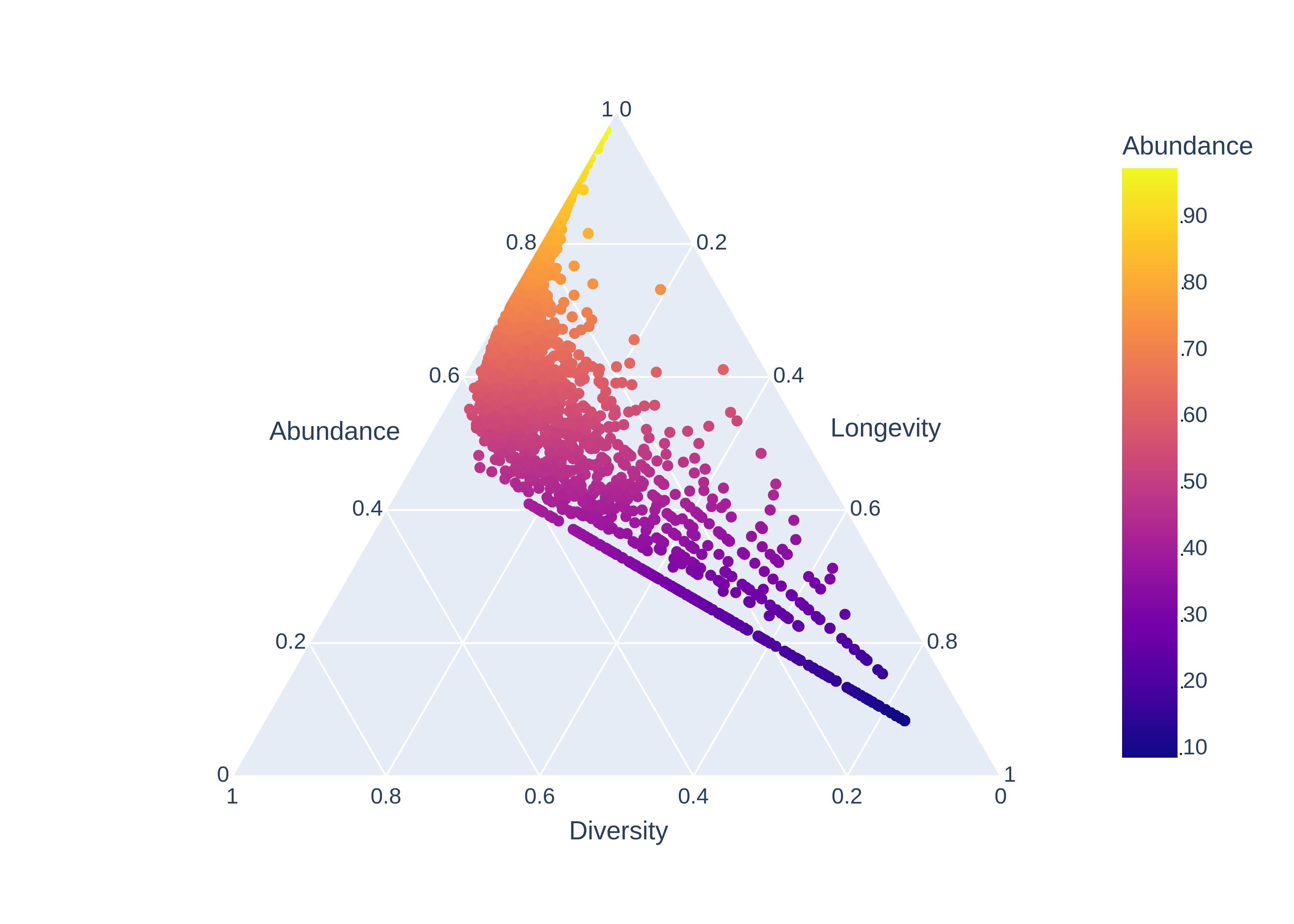} 
    \caption{Ternary Diagram of diversity-longevity and abundance for DENUE 2024. Each point corresponds to a hexagon on the Mexico City map. Hexagons with a large abundance and low diversity tend to have low longevity, as indicated by the lower right triangle corner. As the abundance decreases, longevity and diversity increase. The hexagons' agglomeration is in the 60\% group of each of the three variables. The patterns of straight lines formed by hexagons in the ternary diagram of composition, abundance, and diversity dimensions appear only in peri-urban areas, indicating that these areas are undergoing transitional growth. The dimensions of the three variables (ADL) are very different. To observe the relationship between them, it was necessary to normalize them to one (V/1) for the diagram. For the color legend based on abundance, the normalization was done on 100. Both normalizations can be easily related by considering a 1:10 ratio between the scale and the graph normalization. While abundance decreases, diversity and longevity remain stable until they reach a midpoint that coincides with the midpoint of the normal distribution of longevity (Figure 10). From this point, all three variables decrease similarly, becoming limited to the most heterogeneous hexagons where diversity equals abundance (all economic units are different). This can be identified in the ternary diagram by the straight line that represents the lower limit of the observations.}
    \label{fig:Ternary}
\end{figure}

To further understand the non-linear dynamics previously identified between abundance and diversity, we incorporate longevity (Eq. \ref{eq:equation3}). Figure  \ref{fig:Ternary} shows the interaction among the three structural variables, as a ternary diagram that enables exploration of multivariate interactions under compositional constraints, revealing how shifts in the relative dominance of each component affect the system's scaling behavior. By projecting the observed regimes onto the ternary space, we uncover potential structural transitions and identify regions in which the power-law behavior is either reinforced, attenuated, or replaced by alternative functional forms.

%\begin{equation}
%    y = -167.86 \cdot \log(x) + 0.19 \cdot x + 1141.72
%    \label{eq:mi_formula}
%\end{equation}

%Since the relationships between abundance and diversity, as well as longevity, exhibit non-linear dynamics, power-law and logarithmic models were fitted to clusters of abundance, the results show that the power-law model adequately explains the behavior of the first cluster (low Abundance, with an RMSE of 13.93 and an R² close to 0.68 in 2015, while the logarithmic model is more appropriate for intermediate abundance levels. These models not only improve the explanatory power compared to the linear model, but also allow for the identification of regime transitions in the urban economic structure (from concentration to saturation), which is consistent with theories of self-organization and hierarchical distribution of firms (\parencite{Gabaix2016}; \parencite{Stanley1996}).

The ternary diagram enables us to visualize how longevity influences the relationship between abundance and diversity. The effect of longevity on this relationship may be realized in three ways:
\begin{itemize}
        \item Altering the form of the exponent, modulating the scale in power law behavior.
        \item Acting as a scaling or saturation factor.
        \item Or introducing regime transitions (for example, changing from a power law to a logarithmic function depending on the relative value of the third variable).
\end{itemize}

In this case, diversity has a class limit, unlike abundance, which can have very large values. For this reason, we consider that longevity, given the composition shown in the ternary diagram, has both a modulating effect on the scale factor and a saturating effect on the scale. As for the third option, we consider it less likely, given that the clusters' behavior over the years has been quite consistent.
In Figure \ref{fig:Ternary}, we can observe two straight lines. One at the top of all values (as a limit) and the other in the middle of the values. To understand this behavior, we map the locations of these hexagons. %The locations of these are shown in Figure 10.
As shown in Figure 10, all the hexagons in both straight-line behaviors in the ternary diagram are located in the peri-urban area. Both may reflect a transition phase in the relation between abundance and diversity, as well as between urban and rural areas. The local economy in peri-urban areas is influenced by their links to the city, fulfilling many of the criteria that define the peri-urban \parencite{Donge1992}.
Peri-urban areas represent a transition or interaction zone where urban and rural activities are juxtaposed, and landscape features are subject to rapid changes induced by human activity. These are critical areas of land cover change, often leading to significant transformations.
As the city grows, many peri-urban activities move outward, while others and their associated land uses become integrated into the urban fabric. Peri-urban areas are mosaics of temporary and new residents and activities mingled with long-standing land uses, including farms, villages, quarries, and forest patches \parencite{Douglas2012}.

\section{Discussion}
Among the three properties of economic units in their one-dimensional environment (the Hexagon) that are used as variables in the model, the most informative for assessing resilience—and, more broadly, the economic complexity expressed through these indicators—is Longevity, due to its specific characteristics \parencite{Daepp2015}. Based on the results obtained and consistent with previous findings, Longevity also emerges as the most suitable measure for evaluating the economic relationships observed thus far. In particular, firm survival serves as an effective proxy for identifying more favorable economic environments within the range of observed contexts, as it reflects the interaction among the model’s three dimensions (ADL).
Using the correlation resulting from abundance and diversity to predict longevity, we obtain:

\begin{table}[h!]
\centering
\begin{tabular}{|c|c|c|}
\hline
  & MSE & R² \\ \hline
Genetic & 1.040893 & -0.049480\\ \hline
Linear Regression & 0.984942 & 0.006932\\ \hline
\end{tabular}
\caption{Model performance in predicting longevity using the abundance–diversity correlation as input. Both the genetic and linear regression models exhibit low predictive accuracy, with R² values near zero, indicating that the correlation between abundance and diversity does not explain variations in longevity. The genetic model yields a slightly higher mean square error (MSE = 1.04) and a negative R², suggesting poorer performance than a baseline mean prediction. In contrast, the linear regression model performs marginally better but remains non-informative.}
\end{table}
In both cases (MSE and R²), the model is worse than using the mean. But, when we include the spatial component (using a coordinate system) \footnote{To incorporate spatial relationships among hexagons, we used their planar coordinates $(x, y)$ to represent each observation in a two-dimensional Euclidean space. Based on these coordinates, a $k$-nearest neighbors (kNN) algorithm was implemented to identify, for each hexagon $i$, the set of its $k$ closest spatial neighbors. The Euclidean distance between any two hexagons $i$ and $j$ was calculated as $d_{ij} = \sqrt{(x_i - x_j)^2 + (y_i - y_j)^2}$. For each hexagon, the neighborhood set $N_i$ was defined by the $k$ smallest distances. Once the neighborhood structure was established, we computed the local mean of abundance and diversity among the neighbors of $i$ as $\text{abundance}_{i}^{(\text{neighbors})} = \frac{1}{k} \sum_{j \in N_i} \text{abundance}_j$ and $\text{diversity}_{i}^{(\text{neighbors})} = \frac{1}{k} \sum_{j \in N_i} \text{diversity}_j$. These spatially contextualized variables represent the average abundance and diversity in the immediate surroundings of each hexagon, corresponding to a non-distance-weighted spatial lag model based on a $k$-nearest neighborhood structure.}, the result is the following:

\begin{itemize}
  \item Correlation abundance vs. diversity-neighbors: 0.7081
  \item Correlation diversity vs. abundance-neighbors: 0.7356
\end{itemize}

This output suggests that there is a significant spatial correlation between abundance and diversity in the dataset, regardless of local conditions.
This measures the extent to which the abundance of a location is correlated with the diversity of its neighbors. A value of 0.71 is high, suggesting that local abundance is positively related to the diversity of the environment (neighborhood). It implies that diverse contexts tend to sustain or attract more abundance locally. The value of the correlation between diversity and abundance of neighbours is 0.7357. This measures the extent to which the diversity of a location is correlated with the abundance of its neighbors. This is also a high value (0.74). It implies that local diversity is (positively) influenced by abundant environments.

The influences of urban form and transport infrastructure on economic performance are evident in several contemporary policy debates, notably the "sprawl versus compact city" debate and the future of megacities in the developing world. 

These data illustrate how the behavior and impact of economic units at different scales influence basic urban needs—such as commuting to work, accessing resources, or participating in social activities—within complex urban systems. Typically, work-related trips originate from residential locations. At the same time, cities follow organizational structures that can be positioned along a spectrum from monocentric to polycentric forms. Figures 12 and 13 schematically depict typical mobility patterns in monocentric cities, which often shape residents' movement dynamics.

The location of the residence relative to the central business district (CBD) \parencite{Xu2023} is a decisive factor in determining the mobility scale. In a monocentric city, people with homes closer to the central business district (CBD) typically have more access to various resources, so they travel shorter distances than those living in the periphery, spending less time and money, which increases inequality.

The distribution of people, opportunities, and infrastructure networks is uneven, as seen in the segregation of socioeconomic levels driven by the unequal distribution of facilities and services. This inequality leads to various explicit and subtle disparities in how different individuals can access urban resources like education, employment, and healthcare, which is reflected in urban mobility behavior.

%In this context, the "15-minute city" concept contrasts monocentric cities, where resources are concentrated in a single central area, with polycentric cities that aim to decentralize opportunities and make services accessible within a 15-minute walk or bike ride.

Quantifying urban inequality through the relationship between regional economic performance and mobility behavior data is an important starting point for designing fair and inclusive urban policies that promote both accessibility and equity. Employment distribution plays a relevant role in this dynamic \parencite{Xu2025}. The relationship between urban spatial structure and economic
performance can be best captured at the metropolitan scale, since this scale closely corresponds to the spatial units of labor markets \parencite{Cervero2001}.

%Patterns of human mobility can characterize urban spatial structure. One key determinant of mobility scale is the home location in relation to the central business district (CBD). Building on this insight, \parencite{Xu2023} introduce a mobility centrality index, $\Delta KS$, which captures the variation in commuters' home-centered gyration radius (Rg) as a function of distance from the CBD. This index quantifies the statistical divergence in Rg across different residential groups using the Kolmogorov–Smirnov (KS) test.

%Urban segregation naturally leads to a divergence in mobility behavior. This divergence arises from people's residential locations and the low availability of jobs and facilities in peripheral regions.

\section{Conclusion}
In diverse studies on cities, density and diversity (of people, housing, businesses, and even behavioral patterns) have been considered fundamental variables in their methodology due to their role in determining the city's viability \parencite{Fan2023}.

In this analysis, we suggest that the relationship between economic unit abundance and sectoral diversity may serve as a proxy for systemic longevity, with its sign (positive or negative) depending on the spatial and temporal configuration of the observed data (see the bivariate map, Figure \ref{figure:bivariate}).
On the other hand, the influence of abundance on the diversity of its neighbors, and its diversity on the abundance of its neighbors, generates dynamics that could lead to urban expansion, as is the case with the influence of new public transportation routes or stops. The influences of urban form and transport infrastructure on economic performance are evident in several contemporary policy debates, notably 'sprawl versus compact city' and, in the developing world, the future of megacities \parencite{Cervero2001}.

The generation of spatial behavior toward site choice is based on spatial opportunities. Spatial behavior collectively shapes distribution and composition. Collective distribution is defined by the four types of interaction (competition, cooperation, conflict, and negotiation).
The results presented in the article support the notion that Mexico City is undergoing regionalization. Although it remains a monocentric city, patterns observed in peri-urban areas and economic interactions with neighbors suggest a transition toward a polycentric and regional structure. This process is characteristic of megacities that, due to their size and complexity, begin to functionally integrate their surrounding areas, forming broader urban-regional systems.
Theories such as agglomeration economies \parencite{Cottineau2019} and human mobility \parencite{Xu2025} are important contributions, useful as complements, and are assembled comprehensively in our work on understanding a city as a complex system based on its relations, evolution, behavior, and possible resilience.

The presence of a large informal sector is a defining characteristic of most developing economies. Given its significance, the economic literature on informality has largely focused on its causes and consequences for development. Although less emphasized, the intensive margin accounts for a substantial share of informal employment in developing countries—56 percent in Mexico \parencite{Parra1992}. One of the most significant gaps in current understanding of informal firms is the lack of longitudinal data on their life-cycle behavior. In this context, our findings establish a spatial relationship between abundance and diversity—using longevity data as a proxy for life-cycle dynamics—that can contribute to a more comprehensive depiction of informality in developing countries.

Therefore, economic informality is a crucial issue in developing economies like Mexico. Delving deeper into its dynamics and relationship with the variables studied could provide valuable insights for the design of public policies.

Another key finding is the importance of large companies within the economic system, despite being three orders of magnitude fewer than small and micro-enterprises, as reflected in the power-law distribution. Moreover, by analyzing the relationship between abundance and diversity, it is possible to infer whether the underlying dynamic is one of cooperation or competition, raising the question: Does this dynamic foster informality? Can large economic units (e.g., poles) generate polycentrism and, therefore, be used to create a regionalization process for Mexico's megacities as a public policy that promotes development?

In the case of Mexico City, the interaction between a small number of large economic units and a vastly greater, highly hierarchical center and an extensive network of microeconomic units in the two clusters underscores the central role of this relationship in sustaining informal economic practices and, more broadly, in maintaining the resilience and continuity of the urban economic system.
This duality between an intensive, highly hierarchical center and an extensive, more uniform periphery reflects an urban system characterized by persistent spatial asymmetries. The coexistence of both regimes suggests a model of urban development that remains strongly centralized but is transitioning toward a possible emerging polycentrality, whose stability will depend on the functional articulation between these spaces \parencite{Zhang2017}.

Therefore, in general, the focus must be on developing morphologically functional polycenters in urban areas, strengthening the construction of transportation and other infrastructure to reduce the cost of mobility and the flow of goods and services, creating complementary functions and effective systems of collaboration rather than competition, with a spatial structure that achieves sustainable urban modernization.
Finally, from a complex systems perspective, a more nuanced interpretation could be achieved by modeling these interactions within a heterogeneity–homogeneity framework informed by relational typologies—such as competition, cooperation, and conflict—allowing for the identification of contextual conditions under which each interaction contributes to urban system resilience or fragility.
\subsection{Future work}
\begin{itemize}

\item \textbf{Business demography.} New economic units registered vs. those that disappeared from the directory each year reveal an oscillation of almost 30\% between both flows. We work with this oscillation 
to address the perspective of self-organization as a generator of criticality, as in other studies \parencite{Krafta2020} with the oscillation between densification and expansion. Additionally, we aim to run this model with other cities in Mexico and Europe.
\item \textbf{Deepen the analysis of structural patterns.} Although the model's explanatory power is limited ($R^2 < 0.01$), the primary aim remains to identify structural patterns rather than to generate precise individual predictions. This approach is common in studies of urban complexity.
\item \textbf{Compare morphologies across clusters.} As shown in Figure 7, the graph on the left corresponds to Cluster 1 (0–1000) and the graph on the right to Cluster 2. In both cases, the overall form is consistent despite differences in scale.
\item \textbf{Investigate the role of dominant economic units.} Given the relevance of the small number of large economic units that shape the entire city and generate a monocentric pattern, we propose to examine this phenomenon as a potential gravitational attraction for urban development.
\end{itemize}

\subsection{Limitations}
A recurring concern in the study of the relationship between urbanization, agglomeration, and economic development is the question of reverse causality. Does a country's urban structure drive economic growth, or does economic growth drive urban concentration? The reality is likely to be a combination of both factors \parencite{Frick2018}.
The results of the most recent empirical studies corroborate greater heterogeneity in the relationship between urban concentration and economic growth.
Although the model's explanatory power is limited ($R^2 < 0.01$), the analysis aims to identify structural patterns rather than make precise individual predictions. This strategy is common in studies of urban complexity \parencite{Stanley1996}, \parencite{Hao2021}.

%\section{References}
\printbibliography

%\bibliographystyle{naturemag}
%\bibliography{reference}

%\bibliographystyle{apa}
%\bibliography{reference}

\section{Appendix A -  Economic units by  Size (number of employees) and clusters.}

\begin{figure}[ht]
    \centering
    \includegraphics[width=0.45\textwidth]{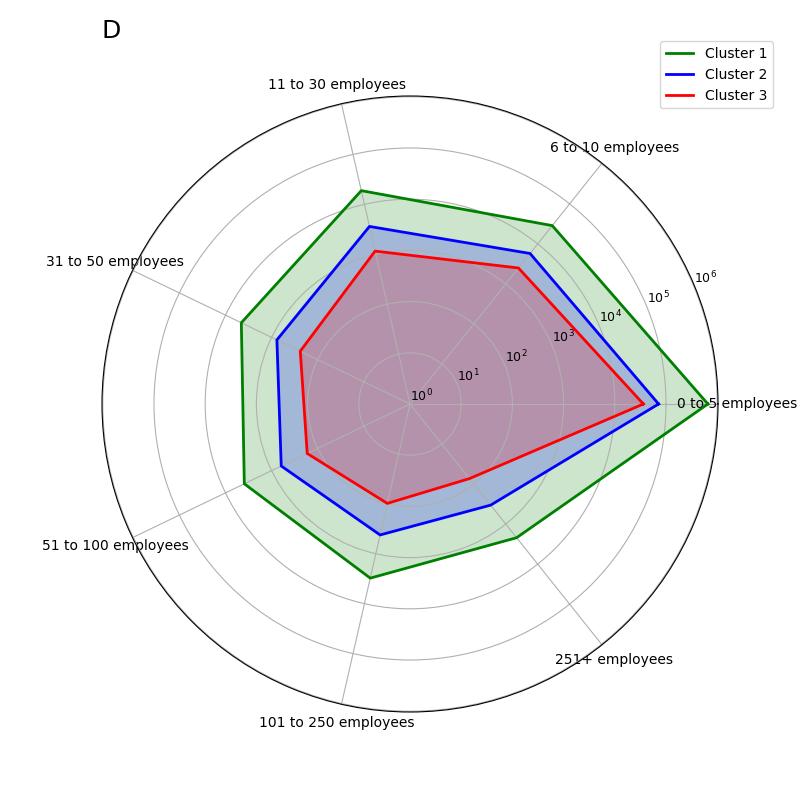}
    \includegraphics[width=0.45\textwidth]{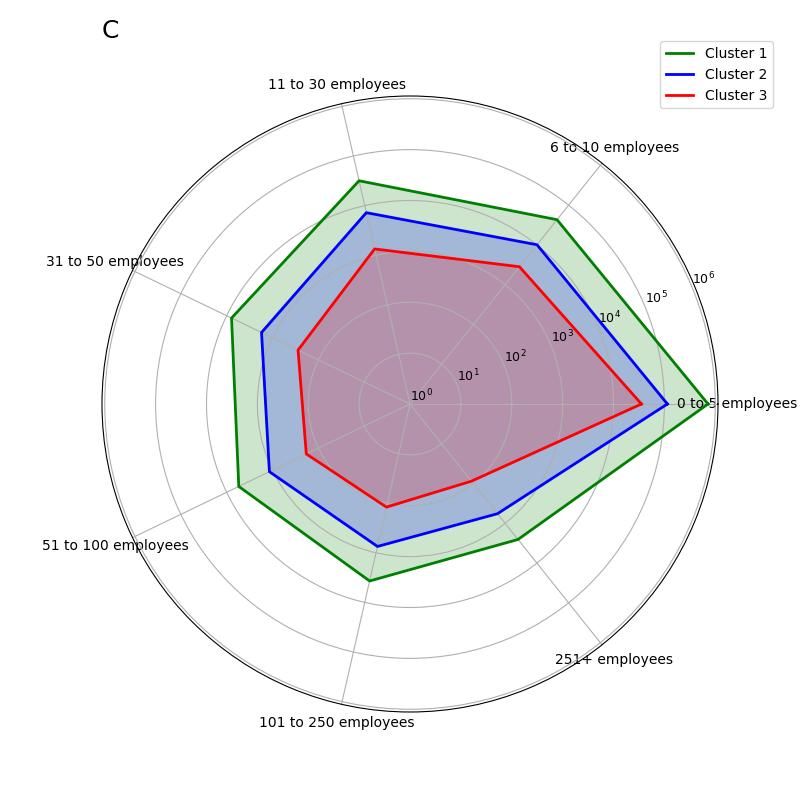}
    
    \caption{Number of economic units by size (number of employees). Although large economic units are orders of magnitude fewer than small and micro economic units, they are structurally significant. These units may function either as early indicators or as cores of emerging processes with qualitative—not merely quantitative—implications, as suggested by power law behavior. This allows us to infer that the relevance of the few large economic units plays a crucial role in shaping urban centrality. Distribution of economic activities based on DENUE 2024. The results indicate a predominance of activities related to services and commerce. Within this group, the most significant categories are (46) Retail Trade, (72) Temporary Accommodation and Food and Beverage Services, and (81) Other Services except Government Activities, which together account for the largest share of establishments. These are followed by (43) Wholesale Trade and (62) Health and Social Assistance Services, which also represent significant components of the urban economic structure. Radar chart with logarithmic scale for all clusters in 2015 and 2024. This relationship between small and large economic units is replicated across all years of the dataset (between 2015 and 2024)}
    \label{fig:radar_employees}
\end{figure}

\section{Appendix B - Data description}

%\begin{figure}[ht]
%    \centering
%    %\includegraphics[width=0.45\textwidth]{Imagenes/distribucion_long_2024.png}
%    \includegraphics[width=0.45\textwidth]{Imagenes/Ab_Long.png}
%    \includegraphics[width=0.45\textwidth]{Imagenes/Div_Long.png}
%    \caption{Longevity distribution (2024) and Longevity correlation with diversity and with Abundance}
%\end{figure}
%\begin{figure}[ht]
%    \centering
%   \includegraphics[width=0.5\textwidth]{Imagenes/Hexagonos_limite.png}
%    \caption{Location Map of the Hexagons form a straight line in the ternary diagram. This behavior occurs because they share the same values for Abundance and diversity, and it is incremental in both longevity and Abundance and diversity. The other straight line in this diagram has this behavior because the diversity value is exactly half the abundance value, and likewise, all three variables are incremental. It is essential to note that these two behaviors occur only in the city's peri-urban area, as indicated on the map.}
%\end{figure}

\begin{table}[h!]
\centering

\begin{minipage}{0.48\textwidth}
\centering
\begin{tabular}{lccc}
\hline
Year & 0 - 1000 & 1001 - 2000 & 2000+ \\ 
\hline
2015 & 3534 & 63 & 9 \\
2016 & 3588 & 96 & 15 \\
2017 & 3589 & 97 & 15 \\
2018 & 3608 & 95 & 14 \\
2019 & 3655 & 105 & 16 \\
2020 & 3703 & 108 & 12 \\
2021 & 3722 & 109 & 13 \\
2022 & 3719 & 109 & 13 \\
2023 & 3749 & 108 & 13 \\
2024 & 3764 & 109 & 12 \\ 
\hline
\end{tabular}
\caption{Number of hexagons by cluster and year}
\end{minipage}
\hfill
\begin{minipage}{0.48\textwidth}
\centering
\begin{tabular}{lccc}
\hline
Year & 0 - 1000 & 1001 - 2000 & 2000+ \\ 
\hline
2015 & 718327 & 83162 & 40350 \\
2016 & 746928 & 120175 & 56161 \\
2017 & 748005 & 121432 & 56190 \\
2018 & 748218 & 119372 & 54000 \\
2019 & 751395 & 130831 & 59352 \\
2020 & 807785 & 134519 & 46182 \\
2021 & 811797 & 135620 & 48622 \\
2022 & 812398 & 135662 & 48620 \\
2023 & 815347 & 134717 & 48573 \\
2024 & 815474 & 136235 & 46464 \\ 
\hline
\end{tabular}
\caption{Sum of economic units by hexagon, cluster, and year}
\end{minipage}

\end{table}
\section{Appendix C - Power law analysis}

We identified the hexagons with abundance values greater than 2000. In 2015, we located 9 hexagons exclusively in the historic center of Mexico City. By 2024, we had located three more hexagons with these characteristics. One is in the new wholesale market, the second is in the Condesa-Chapultepec area, and the third is in the neighboring municipality of Tlanepantla, where an industrial complex and the municipal administrative center are located.
%\begin{figure}[ht]
%    \centering
%    \includegraphics[width=0.8\textwidth]{Imagenes/Abun_plus2000_2015.png}
%    \includegraphics[width=0.8\textwidth]{Imagenes/Abun_plus2000_2024.png}
%    \caption{We located the hexagons with an abundance value greater than 2000. Map A shows the location of the hexagons with an abundance value greater than 2000 economic units for 2015, and Map B shows the location of the hexagons with an abundance greater than 2000 economic units.}
%\end{figure}
%\begin{figure}[ht]
%    \centering
%    \includegraphics[width=0.6\textwidth]{Imagenes/Hexagonos_limite.png}
%    \caption{}
%\end{figure}

\section{Appendix D - Fitting parameters}

\begin{table}[h!]
\centering
\caption{Fitted models for the 0--1000 and 1001--2000 ranges for the years 2015--2024.}
\label{tab:models}
\begin{tabular}{l p{6cm} p{7cm}}
\hline
\textbf{Year} & \textbf{0--1000} & \textbf{1001--2000} \\
\hline
2015 & Power Law: $y = 4.22 \cdot x^{0.55} - 6.99$ & Logarithmic: $y = -167.87 \cdot \log(x) + 0.19 \cdot x + 1141.73$ \\
2016 & Power Law: $y = 4.08 \cdot x^{0.57} - 6.85$ & Logarithmic: $y = 218.39 \cdot \log(x) - 0.09 \cdot x - 1234.38$ \\
2017 & Power Law: $y = 4.10 \cdot x^{0.57} - 6.91$ & Logarithmic: $y = 202.28 \cdot \log(x) - 0.08 \cdot x - 1132.56$ \\
2018 & Power Law: $y = 4.20 \cdot x^{0.56} - 7.06$ & Polynomial: $y = 0.00 \cdot x^{2} + 0.04 \cdot x + 136.86$ \\
2019 & Power Law: $y = 4.07 \cdot x^{0.57} - 6.83$ & Polynomial: $y = -0.00 \cdot x^{2} + 0.23 \cdot x + 24.18$ \\
2020 & Power Law: $y = 4.42 \cdot x^{0.56} - 7.68$ & Logarithmic: $y = 74.13 \cdot \log(x) + 0.01 \cdot x - 334.74$ \\
2021 & Power Law: $y = 4.43 \cdot x^{0.56} - 7.67$ & Logarithmic: $y = 218.09 \cdot \log(x) - 0.10 \cdot x - 1217.35$ \\
2022 & Power Law: $y = 4.44 \cdot x^{0.56} - 7.70$ & Logarithmic: $y = 213.87 \cdot \log(x) - 0.10 \cdot x - 1191.50$ \\
2023 & Power Law: $y = 4.40 \cdot x^{0.56} - 7.53$ & Logarithmic: $y = 251.15 \cdot \log(x) - 0.12 \cdot x - 1423.69$ \\
2024 & Power Law: $y = 4.48 \cdot x^{0.55} - 7.62$ & Logarithmic: $y = 20.54 \cdot \log(x) + 0.05 \cdot x + 1.19$ \\
\hline
\end{tabular}
\end{table}

\begin{figure}[H]
    \centering
    \includegraphics[width=0.45\textwidth]{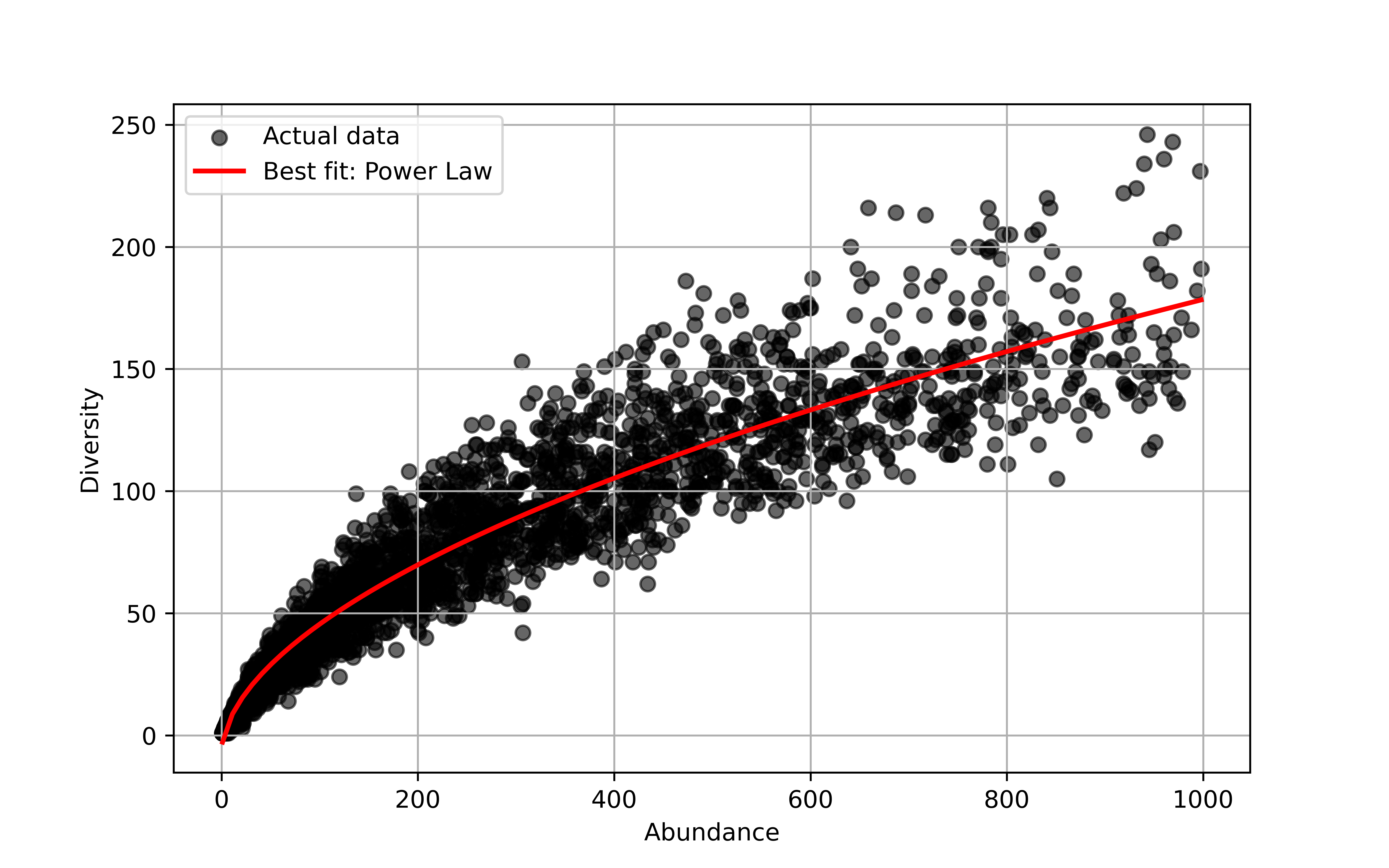}
    \includegraphics[width=0.45\textwidth]{Imagenes/AD_2015_1000-2000.jpg}
    \includegraphics[width=0.45\textwidth]{Imagenes/AD_2024_0-1000.jpg}
    \includegraphics[width=0.45\textwidth]{Imagenes/AD_2024_1000-2000.jpg}
    \caption{Best fitting (power Law) for cluster one (0-1000) and cluster two (1000-2000) in 2015 and 2024. In both cases, a difference in behavior between the two clusters is observed. While cluster 1 exhibits power-law behavior, cluster 2 shows a logarithmic correlation. This behavior is replicated across all years of the dataset (between 2015 and 2024)}
    \label{fig:fittingModels}
\end{figure}

\section{Appendix E - Correlation analysis of ADL Variables: Pearson and Spearman matrices}

\begin{table}[ht]
\centering
\caption{Pearson Correlation Matrix}
\label{tab:pearson-corr}
\begin{tabular}{lccc}
\toprule
 & \textbf{Abundance} & \textbf{Diversity} & \textbf{Longevity} \\
\midrule
\textbf{Abundance} & 1.000000 & 0.822652 & 0.075170 \\
\textbf{Diversity} & 0.822652 & 1.000000 & 0.082190 \\
\textbf{Longevity} & 0.075170 & 0.082190 & 1.000000 \\
\bottomrule
\end{tabular}
\end{table}
\begin{table}[ht]
\centering
\caption{Spearman Correlation Matrix}
\label{tab:spearman-corr}
\begin{tabular}{lccc}
\toprule
 & \textbf{Abundance} & \textbf{Diversity} & \textbf{Longevity} \\
\midrule
\textbf{Abundance} & 1.000000 & 0.985411 & 0.163035 \\
\textbf{Diversity} & 0.985411 & 1.000000 & 0.134074 \\
\textbf{Longevity} & 0.163035 & 0.134074 & 1.000000 \\
\bottomrule
\end{tabular}
\end{table}

\begin{table}[ht]
\centering
\caption{Pairwise Pearson and Spearman Correlations Among Abundance, Diversity, and Longevity}
\label{tab:pairwise-correlations}
\begin{tabular}{lcccc}
\toprule
\textbf{Variable Pair} & \textbf{Pearson $r$} & \textbf{$p$-value} & \textbf{Spearman $\rho$} & \textbf{$p$-value} \\
\midrule
Abundance vs.\ Diversity  & 0.823 & 0.00000 & 0.985 & 0.00000 \\
Abundance vs.\ Longevity  & 0.075 & 0.00000 & 0.163 & 0.00000 \\
Diversity vs.\ Longevity  & 0.082 & 0.00000 & 0.134 & 0.00000 \\
\bottomrule
\end{tabular}
\end{table}

\begin{table}[h!]
\centering
\begin{tabular}{lrrrr}
\hline
\textbf{} & \textbf{n} & \textbf{r} & \textbf{CI95\%} & \textbf{p-value} \\
\hline
Pearson & 3885 & 0.82157 & {[0.81, 0.83]} & 0.0 \\
\hline
\end{tabular}
\caption{Pearson correlation coefficient, sample size, 95\% confidence interval, and p-value.}
\label{tab:pearson-stats}
\end{table} 
The multivariate analysis based on Pearson correlations among \textit{Abundance}, \textit{Diversity}, and \textit{Longevity} reveals a coherent and internally consistent pattern across the three variables. The correlation between Abundance and Diversity exhibits a strong linear association, with a coefficient of $r \approx 0.82$, a narrow 95\% confidence interval ($[0.81,\, 0.83]$), and an effectively zero $p$-value. This indicates a highly stable, statistically significant relationship, confirming that hexagons with higher abundance tend to display greater diversity, and vice versa.

When incorporating \textit{Longevity}, the correlational structure shows that this variable also maintains positive, statistically significant associations with both Abundance and Diversity (with Pearson coefficients consistent with this pattern), though generally with slightly lower magnitudes. This suggests that Longevity is aligned with the local ecological structure, but its linear dependence is less dominant than the relationship observed between Abundance and Diversity. In other words, Longevity appears to capture a more aggregated or temporally extended dimension of the system, whereas Abundance and Diversity reflect more immediate structural characteristics.

Taken together, the three correlations indicate a system in which the variables are not only associated in pairs but also share a coherent statistical architecture. The patterns of abundance and diversity function as partial determinants or predictors of longevity, whereas longevity incorporates temporal information not fully contained in the other two dimensions.

\section{Appendix F - Conceptual Algorithm Description}
%\subsection*{Conceptual Algorithm Description}

\paragraph{Input Data.}
The algorithm starts with a dataset (\texttt{df}) in which each observation represents a spatial unit (in this case, a hexagon). 
Each hexagon has associated variables:\textbf{Abundance}, \textbf{Diversity}, \textbf{Longevity}, \textbf{Spatial coordinates}: \texttt{xcoor}, \texttt{ycoor}

\paragraph{Neighborhood Definition.}
The code uses the $k$-nearest neighbors (kNN) algorithm (\texttt{sklearn.neighbors.NearestNeighbors}) to identify, for each hexagon, the $k$ spatially closest neighbors based on Euclidean distance in the coordinate space. 
In our case, we set $k = 5$.  
Each point's own coordinates are included in the search but later excluded from the neighbor list (\texttt{indices[:, 1:]}).

\paragraph{Spatial Averaging.}
Once the neighbors are identified, the algorithm computes, for each focal hexagon:
\begin{itemize}
    \item the mean abundance of its $k$ nearest neighbors (\texttt{Abundance\_neighbor}), 
    \item the mean diversity of its $k$ nearest neighbors (\texttt{Diversity\_neighbor}).
\end{itemize}
These new variables represent spatially smoothed or locally averaged indicators of the surrounding context.

\subsection*{Spatial Analysis Interpretation}

From a spatial analysis perspective, this procedure performs a \textit{local spatial contextualization}, a common step in spatial statistics, ecology, and geography that evaluates how a spatial unit relates to its surroundings.

\subsubsection*{1. Spatial Weight Matrix (Implicit)}
The kNN model implicitly defines a spatial weights matrix $W$ based on the $k$ nearest neighbors.  
Each hexagon $i$ has five neighbors ($j_1, j_2, \ldots, j_5$), each contributing a uniform weight of $1/k$.

Thus, the computed averages are equivalent to:
\[
A_i^{(\text{neighbors})} = \frac{1}{k} \sum_{j \in N_i} A_j,
\qquad
D_i^{(\text{neighbors})} = \frac{1}{k} \sum_{j \in N_i} D_j,
\]
where $N_i$ is the set of the $k$ nearest neighbors of $i$.

This constitutes a \textit{non-distance-weighted} kNN spatial kernel, meaning all neighbors contribute equally regardless of their exact distance.

\subsubsection*{2. Local Spatial Context}
The resulting variables (\texttt{Abundance\_vecinos}, \texttt{Diversity\_neighbor}) capture the local spatial context:
\begin{itemize}
    \item High values indicate that a hexagon is surrounded by neighbors with high attribute values.
    \item Low values indicate a neighborhood context of low abundance or diversity.
\end{itemize}

These measures are commonly used to:
\begin{itemize}
    \item detect spatial clustering or hotspots,
    \item explore spatial autocorrelation (e.g., Moran's $I$, Geary's $C$),
    \item model spatial spillover effects (e.g., testing whether local diversity influences longevity).
\end{itemize}

\subsubsection*{3. Relation to Spatial Smoothing / Local Indicators}
This operation is conceptually similar to:
\begin{itemize}
    \item a spatial lag model in econometrics (without regression),
    \item a focal mean filter in GIS raster analysis,
    \item kernel smoothing in spatial statistics, using a fixed $k$ rather than a distance bandwidth.
\end{itemize}

\section {Appendix G - Prediction model test.}

\begin{table}[h!]
\centering
\begin{tabular}{lll}
\hline
\textbf{Step} & \textbf{Spatial Concept} & \textbf{Method} \\
\hline
1 & Define space & Coordinates (\texttt{xcoor}, \texttt{ycoor}) define the 2D spatial field \\
2 & Identify local neighborhood & k-nearest neighbors ($k=5$) using Euclidean distance \\
3 & Compute local averages & Mean of variables over neighbors (spatial smoothing) \\
4 & Interpret & Contextualization of each hexagon by local environment \\
\hline
\end{tabular}
\end{table}

%\textbf{Prediction model test.}

A Genetic Programming approach for symbolic regression was implemented using the DEAP library \parencite{Fortin2012}. This evolutionary method enables the discovery of analytical expressions that approximate, in a nonlinear way, the relationship between Abundance and Diversity with respect to Longevity.
The initial population was generated using the half-and-half method (Koza, 1992), and the evolutionary process employed one-point crossover and uniform mutation. A height limit was imposed on the expression trees to prevent excessive growth (bloat) and maintain computational efficiency. The model’s objective function minimized the mean squared error (MSE) between the predicted and observed Longevity values.

\section{Appendix H (DENUE)}
Business Statistical Key (CLEE) Control Data.

This is the unique statistical identification key, assigned exclusively by INEGI to each establishment and company registered in the Mexican Business Statistical Registry. Its function is to identify and link establishments belonging to the same company. This key is mandatory in the administrative records of Economic Units.

It consists of three parts. The first is a static 17-digit code that allows establishments to be identified by their economic activity and size down to the municipal level. The second, for the exclusive use of INEGI, consists of 3 (three) digits that allow INEGI to track changes in the location, economic activity, and size data of Economic Units (as determined by their employed personnel). The third, which is made up of 8 (eight) digits, allows linking establishments of the same company through a national sequential number per company, the letter corresponding to the type of economic unit (head office, branch, or sole), and a verification digit defined by INEGI.
ID or DENUE Identification Number: a 10-digit numerical identification code for each record in the National Statistical Directory of Economic Units (DENUE) database.

\end{document}